# Efficient and Robust Geocasting Protocols for Sensor Networks

Karim Seada, Ahmed Helmy
Electrical Engineering Department, University of Southern California
{seada, helmy}@usc.edu

*Abstract-* **Geocasting is the delivery of packets to nodes within a certain geographic area. For many applications in wireless ad hoc and sensor networks, geocasting is an important and frequent communication service. The challenging problem in geocasting is distributing the packets to all the nodes within the geocast region with high probability but with low overhead. According to our study we notice a clear tradeoff between the proportion of nodes in the geocast region that receive the packet and the overhead incurred by the geocast packet especially at low densities and irregular distributions. We present two novel protocols for geocasting that achieve high delivery rate and low overhead by utilizing the local location information of nodes to combine geographic routing mechanisms with region flooding. We show that the first protocol Geographic-Forwarding-Geocast (GFG) has close-to-minimum overhead in dense networks and that the second protocol Geographic-Forwarding-Perimeter-Geocast (GFPG) provides guaranteed delivery without global flooding or global network information even at low densities and with the existence of region gaps or obstacles. An adaptive version of the second protocol (GFPG\*) has the desirable property of perfect delivery at all densities and close-to-minimum overhead at high densities. We evaluate our mechanisms and compare them using simulation to other proposed geocasting mechanisms. The results show the significant improvement in delivery rate (up to 63% higher delivery percentage in low density networks) and reduction in overhead (up to 80% reduction) achieved by our mechanisms. We hope for our protocols to become building block mechanisms for dependable sensor network architectures that require robust efficient geocast services.** [*]

---

[*] This work was supported by grants from NSF CAREER, Intel and Pratt&Whitney.

# I. INTRODUCTION

Geocasting, transmission of packets to nodes within a certain geographic area, is becoming a crucial communication primitive for many applications in wireless sensor networks. Geocasting could be used to assign tasks to nodes or to query nodes in a certain area. For example, a user may request all sensors in an area where a fire is spreading to report their temperature. Geocasting could also facilitate location-based services by announcing a service in a certain region or sending an emergency warning to a region.

In dependable sensor networks, robust geocasting mechanisms may be necessary for the correct operations of many applications that need the packet to be delivered to all nodes within a region. Sensor networks are expected to be deployed in a wide range of environments, including very harsh environments, therefore robust protocols should be able to cope with different conditions, such as irregular node distributions, gaps and obstacles. As we will show, many of the current geocasting mechanisms become unreliable under these conditions, and robust geocasting mechanisms that consider these environments need to be developed. By robustness here, we mean protocols that are able to reach a maximum number of nodes in the region, while keeping the overhead low in order to conserve energy, which is a critical requirement for sensor network applications. In this work, we develop robust and efficient geocasting mechanisms suitable for different kinds of environments; these protocols provide a high-level of dependability in sensor networks.

In order to preserve the scarce bandwidth and energy consumption of sensor nodes and increase their lifetime, it is desirable to have efficient geocasting mechanisms with low overhead that are able to deliver the data to all nodes within the geocast region. The challenge is that in order to reach all nodes in the region, the packet may have to traverse other nodes outside the region causing extra overhead. There is a tradeoff between the ratio of region nodes reached and the overall overhead incurred due to a geocast transmission. For example, in order to guarantee that all nodes in the region receive a geocast packet, global flooding, by sending the packet to all nodes in the network, may be used which causes very high bandwidth and energy consumption, and can significantly reduce the network lifetime. Other proposed geocast mechanisms that do not rely on global flooding or global information about the network (e.g. [9]) use restricted forwarding zones to limit the number of nodes that forward the geocast packet, and thus they do not guarantee that all nodes in the region receive the packet. This is more significant in sparse networks and networks with irregular distributions or obstacles, where due to disconnections in geographical regions it may not be possible to reach all nodes in the geocast region through a limited forwarding zone. In this paper we will present a mechanism that achieves guaranteed delivery without global flooding and without nodes having global information about the network.

We utilize the geographic location information of nodes to provide two efficient geocast mechanisms with high delivery. Location-awareness is essential for many wireless network applications including geocasting applications, so it is expected that wireless nodes will be equipped with localization techniques that are either based on an infrastructure (e.g. GPS) or ad-hoc based [6]. Geographic routing [8][12] has already shown that utilizing this location information can provide very efficient routing protocols. In this work we extend the benefits of geographic routing to geocast applications.

The first mechanism we propose, Geographic-Forwarding-Geocast (GFG), has close-to-minimum overhead by combining geographic forwarding with region flooding and is ideal in dense networks or in applications where it is sufficient to reach only a proportion of the nodes and guaranteed delivery is not critical. The second mechanism, Geographic-Forwarding-Perimeter-Geocast (GFPG), provides guaranteed delivery to all nodes in the region without global flooding or global information. An adaptive version of the second mechanism is presented which has the desirable property of perfect delivery at all densities in addition to low overhead in dense networks. Extensive simulations that evaluate our mechanisms show the significant improvements provided.

These mechanisms could be used as building blocks for supporting other architectures that require robust geocast services. For example, one of our objectives for building a reliable geocasting mechanism was to provide consistent storage and retrieval of information in *Rendezvous Regions* [17]. Rendezvous Regions is a geographic rendezvous architecture for resource discovery and data-centric storage in large-scale wireless networks. In Rendezvous Regions the network topology is divided into geographical regions, where each region is responsible for a set of keys representing the services or data of interest. Each key is mapped to a region based on a hash-table-like mapping scheme. A few elected nodes inside each region are responsible for maintaining the mapped information. The service or data provider stores the information in the corresponding region and the seekers retrieve it from there. For insertions, we use geocasting to store the information at all the elected servers and for lookups we use anycasting to retrieve the information from any of the servers. In order to achieve consistency between insertions and lookups, we need a geocasting mechanism that can reach all nodes in the region, otherwise lookups may query servers that are not reached by the insertion geocast. GFPG is a perfect match for providing the geocasting component in this architecture.

*Following is a summary of our contributions in this paper:*

*- The design and evaluation of efficient and robust geocasting protocols that combine geographic routing mechanisms with region flooding to achieve high delivery rate and low overhead.*

*- Presenting a guaranteed delivery mechanism based on the observation that by traversing all faces intersecting a region in a connected planar graph, every node of the graph inside the region is traversed. Although this theorem is known, the design of a distributed algorithm that practically and efficiently achieves that in a wireless network is new. Our algorithm is efficient by using a combination of face routing and region flooding, and initiating the face routing only at specific nodes.*

*- Providing an adaptive mechanism in which nodes perform face routing selectively and only when needed based on the density and node distribution in their neighborhood to reduce the unnecessary overhead.*

*- Thorough analysis and comparison of the performance of a class of geocasting protocols under different scenarios.*

The rest of the paper is structured as follows. In Section II we show the related work and previously proposed mechanisms. In Section III we describe the proposed algorithms in detail. In Section IV we evaluate the performance of our mechanisms and compare to previous mechanisms. Section V contains the conclusions.

## II. RELATED WORK

In global flooding, the sender broadcasts the packet to its neighbors, and each neighbor, that has not received the packet before, broadcasts it to its neighbor, and so on, until the packet is received by all reachable nodes including the geocast region nodes. It is simple but has a very high overhead and is not scalable to large networks.

Imielinski and Navas [7][14] presented geocasting for the Internet by integrating geographic coordinates into IP and sending the packet to all nodes within a geographic area. They presented a hierarchy of geographically-aware routers that can route packets geographically and use IP tunnels to route through areas not supporting geographic routing. Each router covers a certain geographic area called a service area. When a router receives a packet with a geocast region within its service area, it forwards the packet to its children nodes (routers or hosts) that cover or are within this geocast region. If the geocast region does not intersect with the router service area, the router forwards the packet to its parent. If the geocast region and the service area intersect, the router forwards to its children that cover the intersected part and also to its parent.

Ko and Vaidya [9] proposed geocasting algorithms to reduce the overhead, compared to global flooding, by restricting the forwarding zone for geocast packets. Nodes within the forwarding zone forward the geocast packet by broadcasting it to their neighbors and nodes outside the forwarding zone discard it. Each node has a localization mechanism to detect its location and to decide when it receives a packet, whether it is in the forwarding zone or not. In the evaluations section we evaluate these algorithms in detail. The algorithms are the following:

- Fixed Rectangular Forwarding Zone (FRFZ) (Figure 1): The forwarding zone is the smallest rectangle that includes the sender and the geocast region. Nodes inside the forwarding zone forward the packet to all neighbors and nodes outside the zone discard it.

- Adaptive Rectangular Forwarding Zone (ARFZ) (Figure 2): Intermediate nodes adapt the forwarding zone to be the smallest rectangle including the intermediate node and the geocast region. The forwarding zones observed by different nodes can be different depending on the intermediate node from which a node receives the geocast packet.

- Progressively Closer Nodes (PCN) (Figure 3): When node B receives a packet from node A, it forwards the packet to its neighbors only if it is closer to the geocast region (center of region) than A or if it is inside the geocast region. Notice that this is different from geographic forwarding; in geographic forwarding a node forwards the packet to the neighbor closest to the region while here a node forwards the packet to all neighbors and *all* neighbors closer to the region forward it further.

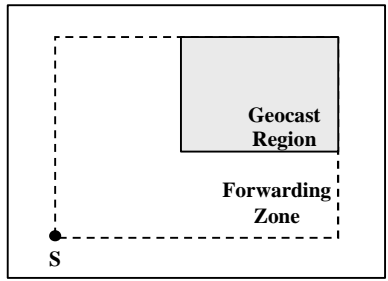

**Figure 1: Fixed Rectangular Forwarding Zone (FRFZ)**

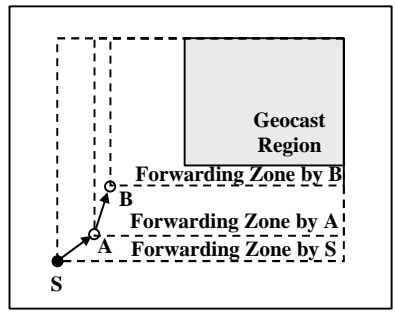

**Figure 2: Adaptive Rectangular Forwarding Zone (ARFZ)**

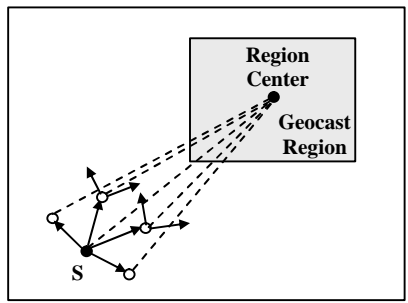

**Figure 3: Progressively Closer Nodes (PCN): Closer nodes to the region than the forwarding node forward the packet further and other nodes discard it**

Other variations of the FRFZ, ARFZ and PCN mechanisms could also be used, for example by increasing the area of the forwarding zone to include more nodes around the geocast region. These variations could improve the delivery rate at the expense of higher overhead, but they do not provide guaranteed delivery. To reduce the overhead further, GeoTORA [10] uses a unicast routing protocol (TORA [16]) to deliver the packet to the region and then floods within the region. Our algorithms also use unicasting to deliver the packet to the region, but we use geographic routing instead of ad-hoc routing protocols. Geographic routing has several advantages: the state kept is minimum, nodes require only information from their direct neighbors so discovery floods and state propagation are not required, and accordingly it has lower overhead and faster response to dynamics. Geographic routing is more scalable than ad hoc routing protocols and more suitable for sensor networks in which the location information is obtained, and since in geocasting, nodes are expected to be aware of their locations anyway, there are no extra costs for using geographic routing.

Variations of global flooding and restricted flooding were presented that use some form of clustering or network divisions to divide the nodes [1][13], such that a single node only in each cluster or division needs to participate in the flooding. This approach can reduce the geocasting overhead by avoiding unnecessary flooding to all nodes at the cost of building and maintaining the clusters. Some approaches (e.g. mesh-based) [2][4] use flooding or restricted flooding only initially, to discover paths to nodes in the geocast region, then these paths are used to forward the packets.

In [20], the network is partitioned using the Voronoi diagram concept and each node forwards the packet to the neighbors whose Voronoi partitions (as seen by the forwarding node) intersect with the geocast region. The idea is to forward to a neighbor only if it is the closest neighbor to any point in the region. Bose *et al.* [3] presented graph algorithms for extracting planar graphs and for face routing in the planar graphs to guarantee delivery for unicasting, broadcasting, and geocasting. For geocasting they provided an algorithm for enumerating all faces, edges, and vertices of a connected planar graph intersecting a region. The algorithm is a depth-first traversal of the face tree and works by defining a total order on the edges of the graph and traversing these edges. An entry edge, where a new face in the tree is entered, needs to be defined for each face based on a certain criteria. In order to determine the entry edges of faces using only local information and without a preprocessing phase, at each edge the other face containing the edge will need to be traversed to compare its edges with the current edge. This could lead to very high overhead. In this paper, we present efficient and practical geocasting protocols that combine geographic routing mechanisms with region flooding to achieve high delivery rate and low overhead.

## A. Geographic Routing

We use geographic routing to efficiently deliver the geocast packet to the region. In addition, our guaranteed delivery algorithm is based on geographic face (also called perimeter) routing. Therefore we provide next a brief overview about geographic routing protocols. Geographic routing consists of greedy forwarding, where nodes move the packet closer to the destination at each hop by forwarding to the neighbor closest to the destination. Greedy forwarding fails when reaching a dead-end (local maximum), a node that has no neighbors closer to the destination. CompassII [11] presented a face routing algorithm that guarantees unicast message delivery on a geometric graph by traversing the edges of planar faces intersecting the line between the source and the destination.

Bose *et al.* [3] presented algorithms and proofs for extracting planar graphs from unit graphs and for face routing in the planar graphs to guarantee delivery. Due to the inefficient paths resulting from face routing, they proposed combining face routing with greedy forwarding to improve the path length. Face routing is used when greedy forwarding fails until a node closer to the destination is reached, then greedy forwarding could be resumed again. GPSR [8] is a geographic routing protocol for wireless networks that works in two modes: greedy mode and perimeter mode. In greedy mode each node forwards the packet to the neighbor closest to the destination. When greedy forwarding is not possible, the packet switches to perimeter mode, where perimeter routing (face routing) is used to route around dead-ends until closer nodes to the destination are found. In perimeter mode a packet is forwarded using the right-hand rule in a planar embedding of the network. Since wireless network connectivity in general is non-planar, each node runs a local planarization algorithm such as GG [5] or RNG [21] to create a planar graph by using only a subset of the physical links during perimeter routing.

III. ALGORITHMS

We present two novel algorithms for geocasting in wireless networks. The first algorithm Geographic-Forwarding-Geocast (GFG) has almost optimal minimum overhead and is ideal for dense networks. The second algorithm Geographic-Forwarding-Perimeter-Geocast (GFPG) provides guaranteed delivery[1] in connected networks even at low density or irregular distributions with gaps or obstacles.

*A. Geographic-Forwarding-Geocast (GFG)*

In geocast applications, nodes are expected to be aware of their geographic locations. Geographic-Forwarding-Geocast utilizes this geographic information to forward packets efficiently toward the geocast region. A geographic routing protocol consisting of greedy forwarding with perimeter (face) routing such as GPSR is used by nodes outside the region to guarantee the forwarding of the packet to the region[2]. Nodes inside the region broadcast the packet to flood the region. An example is shown in Figure 4. In more detail, a node wishing to send a geocast creates a packet and puts the coordinates of the region in the packet header. Then it forwards the packet to the neighbor closest to the destination. The destination of geographic routing in this case is the region center. Each node successively forwards the packet to the neighbor closest to the destination using greedy forwarding. When greedy forwarding fails, perimeter routing is used to route around dead-ends until closer nodes to the destination are found. Ultimately (in case there are nodes inside the region) the packet will enter the region. The first node to receive the geocast packet inside the region starts flooding the region by broadcasting to all neighbors. Each node inside the region that receives the packet for the first time broadcasts it to its neighbors and nodes outside the region discard the packet. For region flooding, smart flooding approaches [15] could also be used to reduce the overhead.

---

[1] In this paper we mean by guaranteed delivery that the routing algorithm itself is guaranteed to deliver the packet to all nodes in the geocast region when the network is connected. The packet may still be dropped for other reasons such as transmission errors or collisions and accordingly some nodes may not receive the packet. In the 802.11 MAC protocol, unicast packets dropped are retransmitted, but broadcasts are unreliable.

[2] Assuming accurate location information. In [18] we studied the effect of location inaccuracy on geographic routing and provided an efficient fix that can be used here as well.

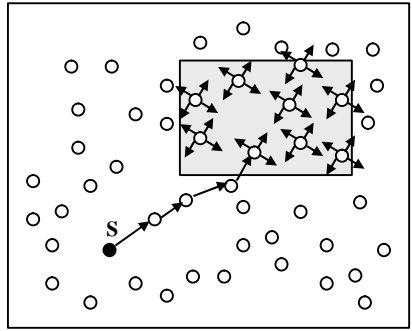

Figure 4: Sender S sends a geocast packet, geographic forwarding is used to deliver the packet to the region, then it is flooded in the region

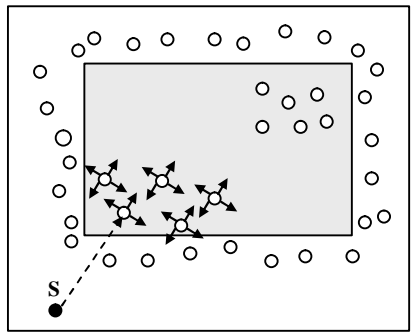

Figure 5: A gap (disconnection) in the geocast region. A packet flooded in the region cannot reach all nodes without going out of the region

In dense networks without obstacles or gaps, GFG is sufficient to deliver the packet to all nodes in the region. In addition, since in dense networks geographic routes are close to optimal routes (shortest path), GFG has almost the minimum overhead a geocast algorithm can have which mainly consists of the lowest number of hops to reach the region plus the number of nodes inside the region itself.

In order for GFG to provide perfect delivery (i.e. all nodes in the region receive the geocast packet), the nodes in the region need to be connected together such that each node can reach all other nodes without going out of the region. In dense networks normally this requirement is satisfied, but in sparse networks or due to obstacles, regions may have gaps such that a path between two nodes inside the region may have to go through other nodes outside the region as shown in Figure 5. In case of region gaps, GFG will fail to provide perfect delivery. The algorithm presented in the next section overcomes this limitation.

## B. Geographic-Forwarding-Perimeter-Geocast

We present an algorithm that guarantees the delivery of a geocast packet to all nodes inside the geocast region, given that the network as a whole is connected. The algorithm solves the region gap problem in sparse networks, but it causes unnecessary overhead in dense networks. Therefore, we present another adaptive version of the algorithm that provides perfect delivery at all densities and keeps the overhead low in dense networks. The adaptive version is not guaranteed as the original version, but the simulation results show that practically it still achieves perfect delivery.

### 1) Guaranteed Delivery (GFPG)

This algorithm uses a mix of geocast and perimeter routing to guarantee the delivery of the geocast packet to all nodes in the region. To illustrate the idea, assume there is a gap between two clusters of nodes inside the region. The nodes around the gap are part of the same planar face. Thus if a packet is sent in perimeter mode by a node on the gap border, it will go around the gap and traverse the nodes on the other side of the gap (see Figure 6 and Figure 8). The idea is to use perimeter routing on the faces intersecting the region border in addition to flooding inside the region to reach all nodes. In geographic face routing protocols as GPSR a planarization algorithm is used to create a planar graph for perimeter routing. Each node runs the planarization algorithm locally to choose the links (neighbors) used for perimeter forwarding. The region is composed of a set of planar faces with some faces totally in the region and other faces intersecting the borders of region as shown in Figure 7. Traversing all faces guarantees reaching all nodes in the region.

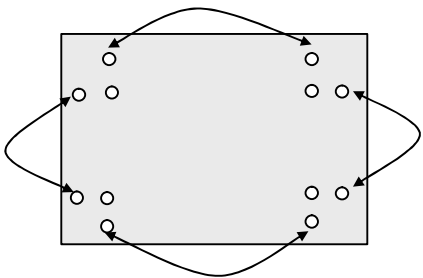

**Figure 6: Perimeter routing connects separated clusters of the same region**

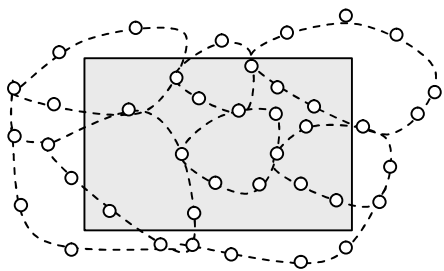

**Figure 7: Planar faces inside and intersecting the region. Traversal of these faces guarantees that every node in the region receives the packet**

We describe now the algorithm in more detail; please refer to Figure 8. Initially, similar to GFG, nodes outside the geocast region use geographic forwarding to forward the packet toward the region. As the packet enters the region, nodes flood it inside the region. All nodes in the region broadcast the packet to their neighbors similar to GFG, in addition, all nodes on the border of the region send perimeter mode packets to their neighbors that are outside of the region. A node is a region border node if it has neighbors outside of the region. By sending perimeter packets to neighbors outside the region (notice that perimeter mode packets are sent only to neighbors in the planar graph not to all physical neighbors), the faces intersecting the region border are traversed. The node outside the region, receiving the perimeter mode packet, forwards the packet using the right-hand rule to its neighbor in the planar graph and that neighbor forwards it to its neighbor and so on. The packet goes around the face until it enters the region again. The first node inside the region to receive the perimeter packet floods it inside the region or ignores it if that packet was already received and flooded before. Notice that all the region border nodes send the perimeter mode packets to their neighbors outside of the region, the first time they receive the packet, whether they receive it through flooding, face routing, or the initial geographic forwarding. This way if the region consists of separated clusters of nodes, a geocast packet will start at one cluster, perimeter routes will connect these clusters together through nodes outside the region, and each cluster will be flooded as the geocast packet enters it for the first time. This guarantees that all nodes in the region receive the packet, since perimeter packets going out of the region will have to enter the region again from the opposite side of the face and accordingly all faces intersecting the region will be covered.

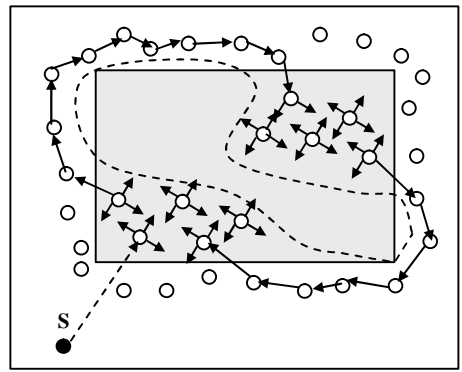

**Figure 8: A mix of region flooding and face routing to reach all nodes in the region. Nodes around the gap are part of the same face. For clarity, here we are showing only the perimeter packet sent around the empty face, but notice that all region border nodes will send perimeter packets to their neighbors that are outside of the region**

### II) Adaptive Algorithm (GFPG*)

Due to the perimeter traversals of faces intersecting the region, the guaranteed algorithm presented in previous section, GFPG, will cause additional overhead that may not be required especially in dense networks, where as we mentioned GFG has optimal overhead by delivering the packet just to nodes inside the region. Ideally we would like perimeter routes to be used only when there are gaps inside the region such that we have perfect delivery also in sparse networks and minimum overhead in dense networks. In this section, we present an adaptation for the algorithm, in which perimeter packets are sent only when there is a suspicion that a gap exists. This new algorithm GFPG*, as we will show in the simulations, practically has perfect delivery in all our simulated scenarios. In this algorithm each node inside the geocast region divides its radio range into four portions as shown in Figure 9(a) and determines the neighbors in each portion. This can be done easily, since each node knows its own location and its neighbors' locations. If a node has at least one neighbor in each portion, it will assume that there is no gap around it, since its neighbors are covering the space beyond its range and so it will not send a perimeter packet and will send only the region flood by broadcasting to its neighbors. If a node has no neighbors in a portion, then it sends a perimeter mode packet using the right-hand rule to the first neighbor counterclockwise from the empty portion as shown in Figure 9(b). Thus the face around the suspected void will be traversed and the nodes on the other side of the void will receive the packet. Notice that in this algorithm there is no specific role for region border nodes and that perimeter packets can be sent by any node in the region, since the gap can exist and need to be detected anywhere. Therefore there are two types of packets in the region, flood packets and perimeter packets. Nodes have to forward perimeter packets even if that packet was flooded before. If a node receives a perimeter packet from the same neighbor for the second time, the packet is discarded, since this means that the corresponding face is already traversed. A node may receive the perimeter packet from different neighbors and thus forwards it on different faces. Figure 10 compares the overhead of GFPG and GFPG* using simulation and shows the improvement achieved by GFPG* in reducing the overhead at high densities. At low densities their overhead is close, since both send the perimeter packets. The details of the simulations are presented in the next section.

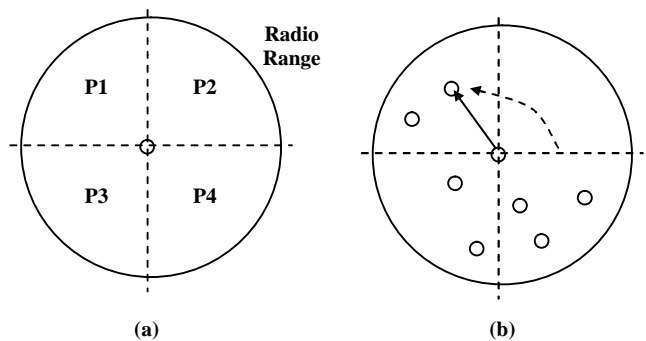

Figure 9: (a) A node divides its radio range into four portions
(b) If a node has no neighbors in a portion, it sends a perimeter packet using the right-hand rule to the first node counterclockwise from the empty portion

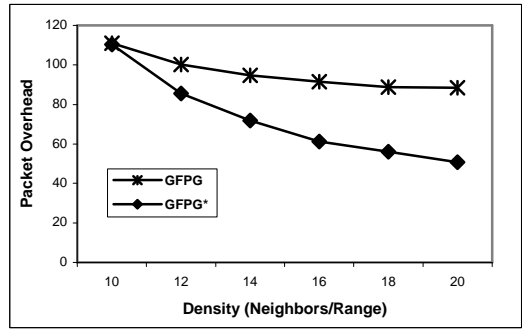

Figure 10: The overhead of the two versions of GFPG. The heuristic added in GFPG* reduces the overhead at high densities while preserving the prefect delivery

GFPG* does not guarantee delivery as GFPG, but our simulation results show that practically it has perfect delivery at all densities, in addition to close-to-minimum overhead at high densities. This is desirable for many types of applications in sensor networks.

IV. PERFORMANCE EVALUATION

In this section we evaluate the performance of GFG and GFPG*, and compare to 4 other geocasting mechanisms [9]:

- Global Flooding

- Fixed Rectangular Forwarding Zone (FRFZ) (Figure 1): The forwarding zone is the smallest rectangle that includes the sender and the geocast region. Nodes inside the forwarding zone forward the packet to all neighbors and nodes outside the zone discard it.

- Adaptive Rectangular Forwarding Zone (ARFZ) (Figure 2): Intermediate nodes adapt the forwarding zone to be the smallest rectangle including the intermediate node and the geocast region. The forwarding zones observed by different nodes can be different depending on the intermediate node from which a node receives the geocast packet.

- Progressively Closer Nodes (PCN) (Figure 3): When node B receives a packet from node A, it forwards the packet only if it is closer to the geocast region (center of region) than A or if it is inside the geocast region.

## A. Main Results

We are interested in evaluating the geocast delivery rate (the ratio of the nodes inside the geocast region that receive the packet) and the geocast overhead (the total number of nodes that forward the geocast packet) of different mechanisms at various densities. In order to have a pure evaluation of the geocast algorithms without interference from other layers such as MAC collisions or physical layer effects, we consider only the routing behavior in an ideal wireless environment of a static 1000-node network. We vary the density of the network by changing the network space area. We present the density as the average number of nodes per radio range. Each simulation run, nodes are distributed at random locations and 10 random senders send a geocast packet to a geocast region in the center of the space. Border regions are studied in Section B. The geocast region size is 1/25 of the space, so it contains an average of 40 nodes. We consider only topologies where the network is connected. The results are computed as the average of 1000 runs. The geographic forwarding protocol used in GFG and GFPG* is GPSR [8] with GG (Gabriel Graph) [5] planarization.

By using random node distributions with different densities we are actually covering a wide range of distributions. At high densities, nodes are more uniformly distributed, while at lower densities (which represents the challenge), the distribution is more irregular and the space contains gaps similar to what obstacles may cause. The random topologies generated have a mix of distributions with some areas uniform and some areas containing gaps of different sizes. The tendency to higher uniformity or gaps depends on the density.

Figure 11 shows the delivery rate of the different geocast mechanisms at densities ranging from an average of 6 neighbors per radio range to 20 neighbors per radio range. Global flooding does not need to be shown, since its delivery rate is always 100% and the overhead is equal to 1000 (the number of nodes) in an ideal wireless environment (notice that we are focusing only on the routing delivery rate and overhead; if the MAC and physical layer effects are included, packets can be dropped for other reasons such as collisions). At high densities all mechanisms have almost perfect delivery. In all mechanisms except GFPG*, the delivery rate decreases at lower densities due to the inability to deliver the packet to all nodes through restricted forwarding zones. GFG and PCN have higher delivery rates than the rectangular forwarding zone mechanisms, which suffer significantly at sparse networks. The reason that GFG has higher delivery than the other mechanisms is that geographic routing (consisting of greedy forwarding and face routing) from the source toward the region is guaranteed to reach the region if the network is connected, while in other mechanisms as FRFZ and ARFZ it may not be possible to reach the region without going out of the forwarding zone. Figure 12 shows the packet overhead. GFG has the lowest overhead since it consists only of the geographic route to the region and the flood inside the region (notice that it is slightly above 40 which is the average number of nodes in the region). GFPG* has a low overhead at high densities, which increases at lower densities to preserve the prefect delivery. PCN, FRFZ and ARFZ have higher overhead at high densities that decrease at low densities accompanied with the reduction in their delivery rate.

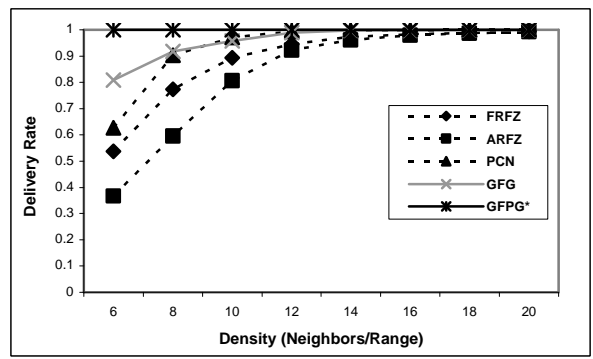

**Figure 11: The delivery rate at different densities**

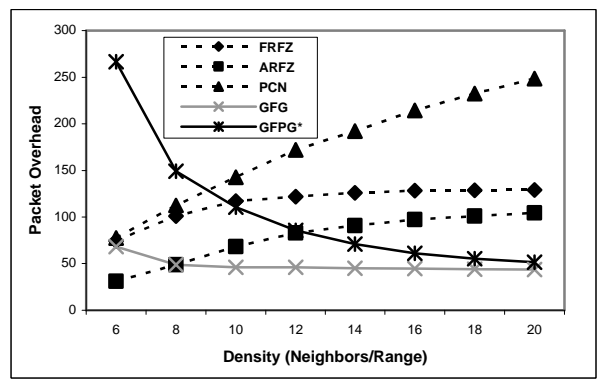

**Figure 12: The packet overhead at different densities**

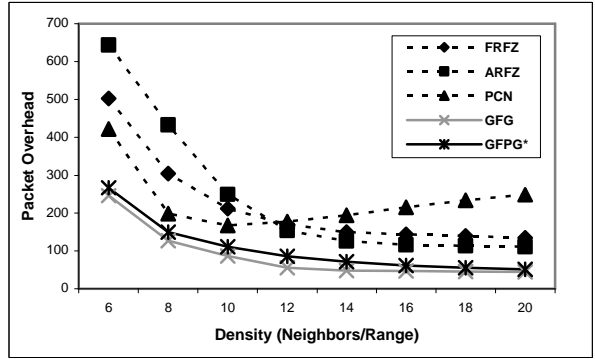

Figure 13: Normalized packet overhead assuming perfect delivery

In order to compare the overheads of protocols without the inverse effect of the delivery rate, we introduce a normalized packet overhead computation assuming that all protocols have a 100% delivery rate by falling back to global flooding for the percentage of delivery that fails[3]. This is only for the sake of analysis (not implemented in the protocols) and to capture the tradeoff between the delivery rate and the overhead. Figure 13 shows the normalized packet overhead. GFG and GFPG* are close with the lowest overhead. PCN has relatively higher overhead at higher densities.

## B. Border Regions

In the previous simulations, the geocast region is close to the center of the network. For regions at the boundary of the network, GFPG* may suffer from long perimeter routes around the external perimeter. In order to avoid the overhead of long perimeter routes, we apply two simple modifications to GFPG*. The first modification is to limit the TTL (we use 10 hops) of the perimeter packet and send it in perimeter mode using both right-hand rule and left-hand rule around the empty portion, such that the packet will not need to go around the whole face and if the face has an opposite side in the region, it will be reached from the shorter direction (an example is shown in Figure 14). The second modification is that nodes close to the boundary do not send the perimeter packets if the empty portion is beyond the network boundary. We run simulations for border regions using these enhancements

---

[3] More exact we compute the normalized packet overhead of a protocol as
 *protocol delivery rate * protocol overhead + (1 - protocol delivery rate) * global flooding overhead*

and as can be seen in Figure 15, Figure 16, and Figure 17 the trends and conclusions are consistent with the previous results.

For the second modification to be applied, nodes need to have approximate knowledge about the network boundaries. While, the first modification of limiting the TTL and sending in both left and right directions could be applied to all nodes independent of their location. If nodes have information about their closeness to the boundary, then it is enough to use this modification only by nodes close to the network boundary.

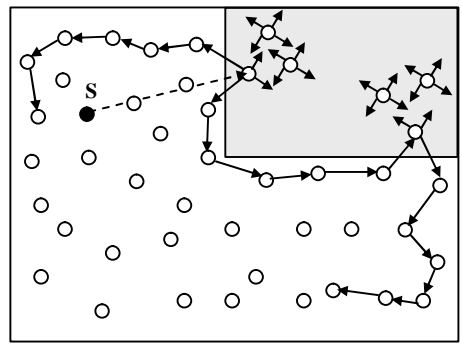

**Figure 14: Sending the perimeter mode packet with a limited TTL using both right-hand rule and left-hand rule. In this figure the TTL is 6**

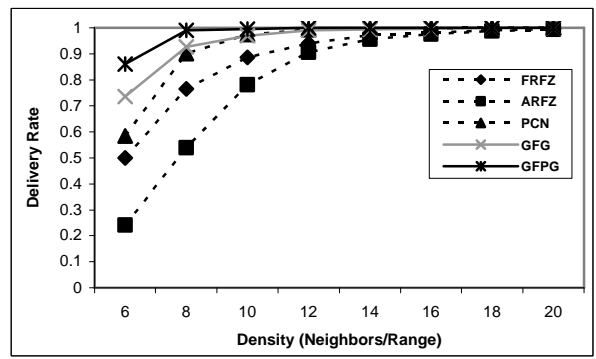

**Figure 15: The delivery rate at different densities with geocast regions close to the border**

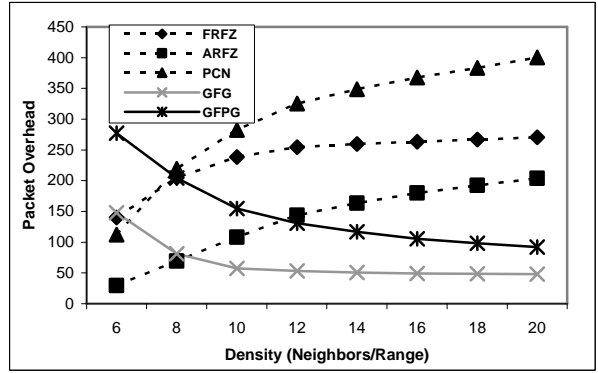

**Figure 16: The packet overhead at different densities with geocast regions close to the border**

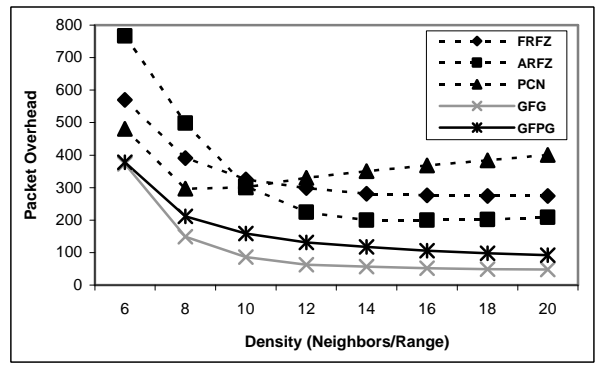

**Figure 17: Normalized packet overhead assuming perfect delivery with geocast regions close to the border**

*C. Summary*

In summary, the results show that GFPG* can have perfect delivery even at very sparse networks without global flooding and at the same time it has close-to-minimum overhead at dense networks. In addition to overcoming gaps, GFPG* could also overcome intermittent breaks in connectivity, which would be a more significant advantage if the wireless physical effects are considered. GFG is a good choice at dense networks and has the lowest overhead. At sparse networks it still has good delivery rate compared to other mechanisms. PCN keeps a good delivery rate, but its overhead is high in dense networks. FRFZ and ARFZ delivery rates decrease fast at lower densities. Other variations of the FRFZ, ARFZ and PCN mechanisms could also be used, for example by increasing the area of the forwarding zone to include more nodes around the geocast region. These variations may improve the delivery rate at the expense of higher overhead, but they still are not adequate for providing perfect delivery.

Another related geocasting algorithm, presented in [3], traverses the planar faces intersecting a region in a certain order based on defining a spanning tree of the faces. The traversal of the nodes inside the region is obtained by traversing the spanning tree. In order to traverse the spanning tree, an entry edge needs to be de determined for each face which represents the link between a *parent* face and a *child* face in the spanning tree. For example, using a depth first traversal, the geocast packet traverses the edges of the initial face (the root of the tree). When it reaches an entry edge, it switches to the new face and starts traversing that face until reaching another entry edge and so on. After finishing traversing a face (and its children) and returning to the entry edge, the packet returns to the parent face and continue on that face. Recursively, all faces of the tree will be traversed. The overhead of the face traversal is comparable to GFPG, but the drawback of this approach is that an extra phase is required for determining the entry edges which contains a significant additional overhead. In order for the geocast packet to be able to identify the entry edges, a preprocessing phase could be used, which traverses all edges in the graph and label them. This complete graph traversal requires flooding the network and will need to be repeated as the topology changes. An online approach for identifying the entry edges would require the geocast packet to check the edges of the opposite face of each edge it traverses. This means that an extra overhead in the order of the number of edges multiplied by the average face size is required. Obviously, this overhead could be very high compared to GFPG which is totally local and does not require any information beyond a single hop.

In the previous simulations, we used a static network, but our mechanisms are also applicable with mobility. Previous studies of geographic routing protocols (e.g. [8]) show fast response to topology changes and higher efficiency with dynamics than non-geographic ad hoc routing protocols. In static networks, the beaconing overhead is negligible, but in mobile networks, more frequent beaconing may be used to detect changes. If a unicast geographic protocol already exists, geocasting will not require additional beaconing overhead than that already incurred by unicast. Otherwise reactive queries for neighbor locations can be used by nodes forwarding packets to the geocast region.

## V. CONCLUSIONS

By exploiting the local geographic information and combining geographic routing mechanisms with region flooding, we presented efficient and robust geocasting mechanisms suitable for dependable sensor networks. We have shown that we can achieve guaranteed delivery without global flooding or global network information by using region flooding and face routing at specific nodes to reach all nodes in the geocast region even with irregular distributions due to gaps or obstacles. The simulations show that our algorithms have significantly lower overhead (up to 80% reduction) than previously proposed mechanisms and that GFPG* has the desirable combination of perfect delivery at all densities and low overhead at high densities. These mechanisms could be used as building blocks for architectures like Rendezvous Regions [17] that require reliable geocasting services.

This is part of our work on assessing and improving the robustness of geographic protocols to non-ideal conditions corresponding to the real-world environments. The conditions considered here are the gaps, obstacles, and irregular distributions with their effect on geocasting. These conditions are common in many sensor networks environments and need to be addressed by robust protocols targeting sensor networks dependability. In other studies, we also examined the effect of lossy links [19] and location inaccuracy [18] on geographic routing, and the effect of mobility and failures on geographic rendezvous mechanisms [17].

## REFERENCES

[1] B. An and S. Papavassiliou. "Geomulticast: Architectures and Protocols for Mobile Ad Hoc Wireless Networks". *Journal of Parallel and Distributed Computing*, v.63 n.2, p.182-195, February 2003.

[2] J. Boleng, T. Camp, and V. Tolety. "Mesh-Based Geocast Routing Protocols in an Ad Hoc Network". IPDPS 2001.

[3] P. Bose, P. Morin, I. Stojmenovic, and J. Urrutia. "Routing with Guaranteed Delivery in Ad Hoc Wireless Networks". *Workshop on Discrete Algorithms and Methods for Mobile Computing and Communications (DialM 1999)*.

[4] T. Camp and Y. Liu. "An Adaptive Mesh-Based Protocol for Geocast Routing". Journal on Parallel and Distributed Computing: Special Issue on Routing in Mobile and Wireless Ad Hoc Networks, Feb. 2003, pp. 196–213.

[5] K. Gabriel and R. Sokal. "A New Statistical Approach to Geographic Variation Analysis". *Systematic Zoology 18* (1969), 259–278.

[6] J. Hightower and G. Borriello. "Location Systems for Ubiquitous Computing". *IEEE Computer,* Aug. 2001.

[7] T. Imielinski and J. Navas. "GPS-based Addressing and Routing". IETF RFC 2009, Nov. 1996.

[8] B. Karp and H. Kung. "GPSR: Greedy Perimeter Stateless Routing for Wireless Networks". *ACM MOBICOM* 2000.

[9] Y. Ko and N. Vaidya. "Flooding-based Geocasting Protocols for Mobile Ad Hoc Networks". *ACM/Baltzer Mobile Networks and Applications (MONET) Journal,* Dec. 2002.

[10] Y. Ko and N. Vaidya. "Anycasting-based Protocol for Geocast service in Mobile Ad Hoc Networks". *Computer Networks Journal,* April 2003.

[11] E. Kranakis, H. Singh, and J. Urrutia. "Compass Routing on Geometric Networks". *In Proc. 11th Canadian Conference on Computational Geometry*, August 1999.

[12] F. Kuhn, R. Wattenhofer, and A. Zollinger. "Worst-Case Optimal and Average-Case Efficient Geometric Ad-Hoc Routing". *ACM MOBIHOC* 2003.

[13] W.H. Liao, Y.C. Tseng, K.L. Lo, and J.P. Sheu. "GeoGRID: A Geocasting Protocol for Mobile Ad Hoc Networks Based on GRID". *Journal of Internet Technology*, vol. 1, no. 2, pp. 23–32, Dec. 2000.

[14] J. Navas and T. Imielinski. "Geographic Addressing and Routing". *ACM MOBICOM* 1997.

[15] S. Ni, Y. Tseng, Y. Chen, and J. Sheu. "The Broadcast Storm Problem in a Mobile Ad Hoc Network". *ACM MOBICOM* 1999.

[16] V. Park and M. Corson. "A Highly Adaptive Distributed Routing Algorithm for Mobile Wireless Networks". *IEEE INFOCOM* 1997.

[17] K. Seada and A. Helmy. "*Rendezvous Regions*: A Scalable Architecture for Service Location and Data-Centric Storage in Large-Scale Wireless Networks". *IEEE/ACM IPDPS 4th International Workshop on Algorithms for Wireless, Mobile, Ad Hoc and Sensor Networks (WMAN), Santa Fe, New Mexico,* April 2004.

[18] K. Seada, A. Helmy, and R. Govindan. "On the Effect of Localization Errors on Geographic Face Routing in Sensor Networks". *IEEE/ACM 3rd International Symposium on Information Processing in Sensor Networks (IPSN), Berkeley, CA,* April 2004.

[19] K. Seada, M. Zuniga, A. Helmy, and B. Krishnamachari. "Energy-Efficient Forwarding Strategies for Geographic Routing in Lossy Wireless Sensor Networks". *ACM SenSys* 2004.

[20] I. Stojmenovic, A. P. Ruhil, and D. K. Lobiyal. "Voronoi Diagram and Convex Hull-Based Geocasting and Routing in Wireless Networks". IEEE ISCC 2001.

[21] G. Toussaint. "The Relative Neighborhood Graph of a Finite Planar Set". *Pattern Recognition 12*, 4 (1980), 261–268.

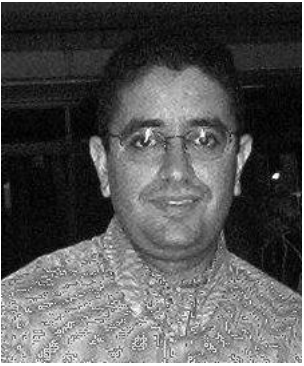
Karim Seada is a Ph.D. candidate in the Electrical Engineering Department at the University of Southern California, Los Angeles. He received his M.S. (2000) and B.S. (1998) with honors in Computer Engineering from Cairo University, Egypt and M.S. (2004) in Computer Science from the University of Southern California. His current research interests lie in the area of computer networks and distributed systems with emphasis on the robustness of geographic protocols in wireless networks, resource discovery in ad hoc and sensor networks, design and testing of network protocols, and multicast congestion control. In summer 2002, he has been an intern at Intel's Network Architecture Lab working in IP mobility. He is a student member of the IEEE and ACM.
URL: http://nile.usc.edu/~seada

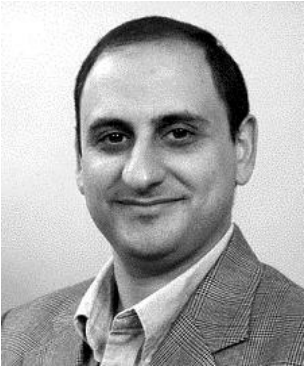
Ahmed Helmy received his Ph.D. in Computer Science (1999), M.S. in Electrical Engineering (1995) from the University of Southern California, M.S. Eng Math (1994) and B.S. in Electronics and Communications Engineering (1992) from Cairo University, Egypt. Since 1999, he has been an Assistant Professor of Electrical Engineering at the University of Southern California. In 2002, he received the National Science Foundation (NSF) CAREER Award. In 2000 he received the USC Zumberge Research Award, and in 2002 he received the best paper award from the IEEE/IFIP International Conference on Management of Multimedia Networks and Services (MMNS). In 2000, he founded -- and is currently directing -- the wireless networking laboratory at USC. His current research interests lie in the areas of protocol design and analysis for mobile ad hoc and sensor networks, mobility modeling, design and testing of multicast protocols, IP micro-mobility, and network simulation.
URL: http://ceng.usc.edu/~helmy